\documentclass[aps,groupedaddress,superscriptaddress,amssymb,amsmath,amsbsy,amsfonts,twocolumn,prl]{revtex4-1}

\usepackage{amssymb,amsmath,amsthm,color,graphicx,times,graphicx}
\usepackage{hyperref}
\usepackage{enumerate}
\usepackage{subfigure}
\usepackage{dsfont}
\usepackage{times}
\usepackage{txfonts}
\usepackage{bbold}

\theoremstyle{plain}

\theoremstyle{definition}

\date{August 30, 2017}

\begin{document}

\title{Optimal Continuous Variable Quantum Teleportation with Limited Resources}

\author{Pietro Liuzzo-Scorpo}
\affiliation{Centre for the Mathematics and Theoretical Physics of Quantum Non-Equilibrium Systems (CQNE), \\ $\mbox{School of Mathematical Sciences, The University of Nottingham,
University Park, Nottingham NG7 2RD, United Kingdom}$}

\author{Andrea Mari}
\affiliation{NEST, $\mbox{Scuola Normale Superiore and Istituto Nanoscienze-CNR, I-56127 Pisa, Italy}$}

\author{Vittorio Giovannetti}
\affiliation{NEST, $\mbox{Scuola Normale Superiore and Istituto Nanoscienze-CNR, I-56127 Pisa, Italy}$}

\author{Gerardo Adesso}
\affiliation{Centre for the Mathematics and Theoretical Physics of Quantum Non-Equilibrium Systems (CQNE), \\ $\mbox{School of Mathematical Sciences, The University of Nottingham,
University Park, Nottingham NG7 2RD, United Kingdom}$}

\begin{abstract}
Given a certain amount of entanglement available as a resource, what is the most efficient way to accomplish a quantum task? We address this question in the relevant case of continuous variable quantum teleportation protocols implemented using two-mode Gaussian states with a limited degree of entanglement and energy. We first characterize the class of single-mode phase-insensitive Gaussian channels that can be simulated via a Braunstein--Kimble protocol with non-unit gain and minimum shared entanglement, showing that infinite energy is not necessary apart from the special cases of the quantum limited attenuator and amplifier. We also find that, apart from the identity, all phase-insensitive Gaussian channels can be simulated through a two-mode squeezed state with finite energy, albeit with a larger entanglement. We then consider the  problem of teleporting single-mode coherent states with Gaussian-distributed displacement in phase space. Performing a geometrical optimization over phase-insensitive Gaussian channels, we determine the maximum average teleportation fidelity achievable with any finite entanglement and for any realistically finite variance of the input distribution.
\end{abstract}

\maketitle

Determining the ultimate performance of quantum technologies in the presence of limited resources is essential to gauge their usefulness in the real world. {\em Quantum  teleportation} \cite{Bennett1993,Vaidman1994} enables the ``disembodied transfer'' of an unknown quantum state from a sender to a remote receiver, usually named  Alice and Bob, respectively. To accomplish this, they need to share a quantum resource, i.e., an entangled state, and to classically communicate. Ideally, if the resource is maximally entangled, Bob can retrieve an exact copy of the input state.
Unfortunately this is unrealistic, especially for continuous variable systems \cite{VanLoock2002,Braunstein2005,Pirandola2006}, where maximal entanglement can be obtained only in the unphysical limit of infinite energy \cite{Vaidman1994,Giedke2002}. It is thus important to identify the most efficient teleportation schemes, which make optimal use of limited quantum resources to achieve the largest {\em fidelity} between input and output, averaged over a  specified set of input states \cite{Braunstein2000}.
For discrete variable systems, with uniformly sampled pure input states, a relation between such optimal fidelity and the entanglement of the resource was found in \cite{Horodecki1999}.
%In general, this is a nontrivial task which involves an optimization over all quantum channels which describe teleportation protocols, subject to additional constraints depending on the input set and the available resource.
This Letter solves such a problem in a prominent continuous variable scenario.

A practical protocol for continuous variable teleportation, which employs a two-mode (finitely) squeezed state as a quantum resource at the price of realizing an imperfect teleportation, was proposed by Braunstein and Kimble (BK) \cite{Braunstein1998} and implemented in several experiments \cite{Furusawa1998,Bowen2003,NP2015}.  A characteristic feature of the BK protocol is that the input and output states are connected by a Gaussian additive noise channel \cite{Ban2002}. Moreover, by simply introducing a non-unit classical gain in the BK protocol \cite{Furusawa1998,Polkinghorne1999,Hofmann2000,Ide2002}, more general effective Gaussian channels can be ``simulated'' by teleportation, including quantum attenuators and amplifiers.
A natural question emerging in this context is the following: Optimizing over all teleportation protocols and general resource states, how much entanglement is necessary to simulate a given Gaussian channel? A simple lower bound is given by the entanglement of the so-called Choi state associated to the channel \cite{Giedke2002,Mari2008, Niset2009,Holevo2011,Pirandola2015}; such a bound is  achievable using resources with infinite mean energy.  As shown in fact in \cite{Giedke2002,Niset2009,Pirandola2015}, any Gaussian channel can be deterministically implemented through a BK protocol exploiting the respective Choi state as a quantum resource.

In the first part of this Letter we focus on single-mode phase-insensitive Gaussian channels, which model typical sources of noise in quantum optics \cite{Caves1994,Mari2014,Giovannetti2014}.  We show that almost all of them (but the quantum limited attenuator and amplifier) can be implemented by teleportation with more realistic resource states, having the same entanglement as the Choi state but finite mean energy. Moreover, through a non-unit gain BK teleportation based on pure two-mode squeezed states (TMSSs) with finite energy yet with larger entanglement, one can simulate all phase-insensitive Gaussian channels but the identity.

In the second part of this Letter we consider the concrete problem of teleporting an alphabet of coherent states \cite{Glauber1963} sampled from a phase-invariant Gaussian distribution with  finite variance. Maximizing over all phase-insensitive Gaussian channels, we determine the optimal average teleportation fidelity achievable as a function of the shared entanglement and input  variance. Our result generalizes several previous studies partially addressing similar questions. For example in \cite{Braunstein2000,Hammerer2005,Namiki2011,Chiribella2014,Yang2014}  the fidelity was maximized over classical strategies (i.e., with zero shared entanglement) identifying the \emph{classical benchmark} for different input sets. For a fixed entanglement, the optimal average fidelity for teleporting  coherent states was studied in \cite{Adesso2005,Mari2008}, albeit assuming an ideal flat distribution with unbounded variance. The optimization of the most realistic scenario given by  a finite input variance remained hitherto unsolved, and is settled by the present Letter.
 %settles this open question, determining optimal and feasible continuous variable teleportation schemes.

\emph{Gaussian states and Gaussian channels} --- An $m$-mode bosonic system \cite{Braunstein2005,Weedbrook2012,Adesso2014,Serafini2017} is usually described in terms of a vector of quadrature operators  $\hat{R}=(\hat{q}_1, \hat{p}_1,\dots,\hat{q}_m,\hat{p}_m )^{\top}$ satisfying the canonical commutation relations $[\hat{R}_j,\hat{R}_k]=i \Omega_{jk}$, with $\Omega=i \sigma_y^{\oplus m}$. Here and in the rest of this Letter, $\mathbb{1},\sigma_x,\sigma_y$, $\sigma_z$ are the $2\times2$ identity and Pauli matrices respectively.

Gaussian quantum states can be defined as Gibbs ensembles of quadratic Hamiltonians and are fully characterized by the first and second statistical moments of the quadrature operators, i.e., the displacement vector $d=\langle\hat{R}\rangle$ and the covariance matrix $V_{jk}=\langle\{\hat{R}_j-d_j,\hat{R}_k-d_k\}_+\rangle$, where $\{\cdot ,\cdot \}_+$ is the anti-commutator.
 In order for a symmetric covariance matrix $V$ to describe a physical state, it has to satisfy the Robertson-Schr\"{o}dinger uncertainty relation  $V \ge i\Omega$ \cite{Simon1994}. For instance, in this notation, single-mode coherent states $|\alpha \rangle$ (with $\alpha \in \mathbb C$) \cite{Glauber1963} are minimum uncertainty Gaussian states specified by $d=\big(\!\sqrt{2}\,\mbox{Re}\,\alpha,\sqrt{2}\,\mbox{Im}\,\alpha\big)^\top$ and $V=\mathbb{1}$.
The entanglement of any two-mode Gaussian state with covariance matrix $V$ can be quantified by the \emph{logarithmic negativity} \cite{Eisert2001,Vidal2002,Adesso2004,Plenio2005} $E_\mathcal{N}=\log || \rho^\Gamma||_1= \max\{-\log\tilde{\nu}_-,0\}$, where $\tilde{\nu}_-$ is the lowest symplectic eigenvalue of the covariance matrix $\tilde V=(\mathbb 1 \oplus \sigma_z) V (\mathbb 1 \oplus \sigma_z)$ associated to the partially transposed state $\rho^\Gamma$. The mean energy of a $m$-mode Gaussian state with zero first moments, i.e.~the expectation value of the non-interacting quadratic Hamiltonian, can be easily computed from the covariance matrix $V$ of the state \cite{Adesso2014}. In units of $\hbar \omega$ this is given by: $\bar{n}=(1/m)\sum_{k=1}^m \langle\hat{a}_k^\dagger\hat{a}_k\rangle_\rho= ( {\rm Tr}~V / m - 2)/4$.

\begin{figure}[t!]
\center
\includegraphics[width=8.0cm]{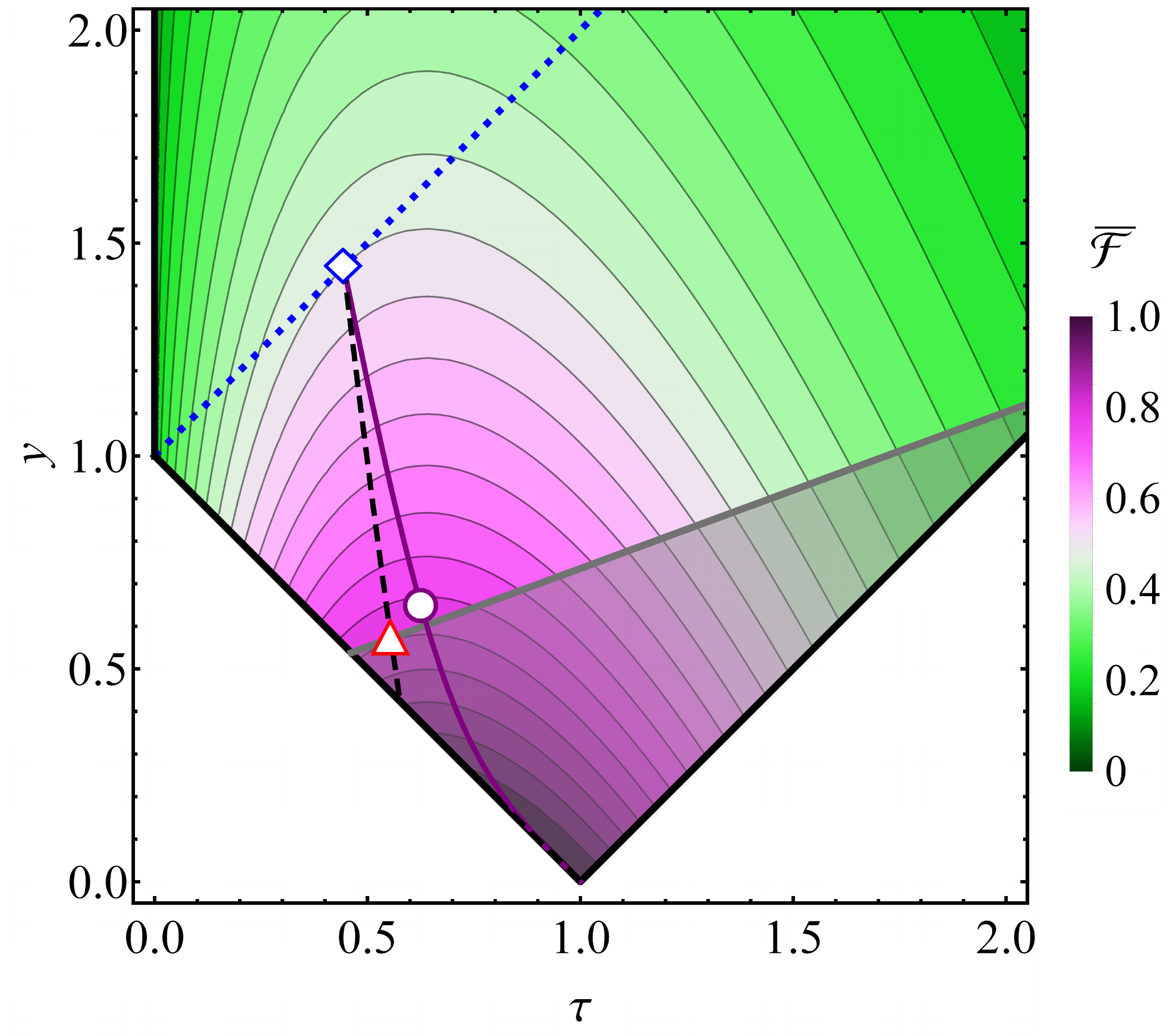}\vspace*{-.2cm}
    \caption{(Color online) Diagram of phase-insensitive single-mode Gaussian channels in the $(\tau,y)$ plane based on the parametrization \eqref{pic}. The white area corresponds to unphysical channels, delimited by \eqref{tauyCPT}.    The shaded gray area corresponds to channels not accessible by teleportation schemes with finite entanglement $E_{\cal N}=2r$, according to the bound \eqref{boundgray}. Channels above the (blue) dotted line are  entanglement-breaking, according to \eqref{tauyEB}. The colored contour plot depicts the average fidelity $\bar{\cal F}$ \eqref{eq:fidelity} associated to each phase-insensitive channel, for an input ensemble of coherent states with phase space variance $\lambda^{-1}$. Relevant channels are highlighted as special points: (triangle) optimal teleportation scheme \eqref{Fopt}; (circle) suboptimal teleportation scheme based on a TMSS \eqref{FTMSS}; (diamond) optimal measure-and-prepare strategy achieving the classical benchmark \eqref{Fclass}. The (black) dashed and (purple) solid lines  represent the channels achievable through the considered teleportation schemes by varying the entanglement parameter  $0\le r < \infty$. The diagram is a snapshot at $r=\lambda=0.5$. All the plotted quantities are dimensionless. \quad  \\[-.4cm]}
\label{fig}
\end{figure}

Gaussian channels are completely positive trace-preserving maps which preserve the Gaussianity of quantum states \cite{Caves1994,Braunstein2005,Wolf2007,Weedbrook2012,Serafini2017}. They can be represented (up to additional displacements) by two matrices $(X,Y)$, with $Y=Y^\top$, which act on the displacement vector and the covariance matrix as
\begin{equation} \label{GChannel}
d\rightarrow Xd, \qquad V\rightarrow XVX^{\top}+Y,
\end{equation}
and satisfy the complete positivity condition $Y+iX\Omega X^{\top}\geq i\Omega$. The latter, for single-mode Gaussian channels, reduces to:
%\begin{equation}\label{eq:physicalregion}
$Y\ge0,\  \sqrt{\det Y}\geq|1-\det X|$.
%\end{equation}
Moreover, single-mode Gaussian channels for which $\sqrt{\det Y}\geq |\det X|+1$ are entanglement-breaking \cite{Holevo2008,Schaefer2013,Mari2014,Giovannetti2014}, i.e.~when acting on one mode of any bipartite system they always produce a separable  output state. These channels correspond to classical measure-and-prepare protocols, hence can be trivially simulated by Alice and Bob via classical communication only. In this Letter we will mainly focus on phase-insensitive single-mode channels defined by:
\begin{equation}
X=\sqrt{\tau} \mathbb{1}, \qquad Y=y \mathbb{1},  \label{pic}
\end{equation}
where $\tau$ and $y$ are scalars representing transmissivity (or gain) and added noise, respectively (in the notation of  \cite{Schaefer2013}).
In the plane $(\tau,y)$ illustrated in Fig.~\ref{fig}, we have thus
\begin{align}
y &\ge |1- \tau|\ \Leftrightarrow \  \mbox{completely positive},  \label{tauyCPT} \\
y &\ge 1+ |\tau| \  \Leftrightarrow \ \mbox{entanglement-breaking} \label{tauyEB}.
\end{align}
We will restrict to $\tau  \geq 0$, which excludes phase-contravariant channels. The channels on the lower boundary of the completely positive region, i.e.~with $y=|1- \tau|$, correspond in particular  to the quantum limited attenuator (also known as pure loss channel) for $0 \leq \tau <1$,  and to the quantum limited amplifier for $\tau >1$, with $\tau=1$ denoting the identity channel.

\emph{Continuous variable teleportation} --- We now briefly review the BK teleportation protocol \cite{Braunstein1998}, and we determine the induced Gaussian channel connecting input and output modes.
Assume that Alice and Bob share a generic quantum resource consisting of a two-mode Gaussian state with zero first moments and covariance matrix
%\begin{equation}\label{eq:resource}
$  V_{AB}=\left( \begin{array}{cc}
    A & C  \\
    C^{\top} & B \end{array}\right)$,
%\end{equation}
where $A$, $B$, and $C$ are $2\times2$ real matrices, and consider an unknown input state with displacement vector $d_{\rm in}$ and covariance matrix $V_{\rm in}$. Alice mixes the input state with her part of the resource state through a balanced beam-splitter and measures, via homodyne detection on each mode, the two commuting quadratures $\hat{Q}_+=(\hat{q}_{\rm in}+\hat{q}_A)/\sqrt{2}$ and $\hat{P}_-=(\hat{p}_{\rm in}-\hat{p}_A)/\sqrt{2}$.   Then, Alice classically communicates her measurement results (photocurrents) to Bob who performs a displacement on his part of the shared state: $\hat{q}_B\rightarrow\hat{q}_{\rm out}=\hat{q}_B+g\sqrt{2}{Q}_+$, $\hat{p}_B\rightarrow\hat{p}_{\rm out}=\hat{p}_B+g\sqrt{2}{P}_-$. Here $g>0$ is the \emph{gain} parameter which is usually set to $g=1$ when teleportation is based on maximally entangled states. However, in general, a non-unit gain has been studied \cite{Furusawa1998,Polkinghorne1999,Ide2002,Hofmann2000} in the context of realistic resources. Once the protocol is completed, the displacement vector and covariance matrix of the output state can be computed using the methods of \cite{Fiurasek2002, Pirandola2006, VanLoock2002,Wolf2007} and are found to be:
\begin{equation}\mbox{$d_{\rm out}= g\, d_{\rm in}$, \quad
 $V_{\rm out}=g^2V_{\rm in}+g^2 \sigma_z A\sigma_z+g(\sigma_z C+C^{\top} \sigma_z)+B$}.
 \end{equation}
These formulas show that the non-unit gain BK protocol induces a Gaussian channel with a diagonal gain matrix $X=g \mathbb 1$, and a noise matrix $Y$ which depends on the covariance matrix $V_{AB}$ of the resource. If we set $V_{AB}$ in block standard form:
\begin{align}\label{normal}
A= a\mathbb 1, \quad B= b \mathbb 1 , \quad C=-c \sigma_z. , \qquad a,b,c \in \mathbb R_+,
\end{align}
then also the noise matrix induced by teleportation is proportional to the identity and the protocol is equivalent to the phase-insensitive channel defined in \eqref{pic}, with parameters
\begin{align}
\tau=g^2, \qquad y= g^2 a -2 g c +b. \label{simul}
\end{align}

\emph{Implementable phase-insensitive channels}  ---
Here we are interested in the inverse problem: we would like to understand, given an arbitrary pair $(\tau,y)$, what covariance matrix $V_{AB}$ can be used as a resource to simulate the corresponding channel with a minimum amount of entanglement (quantified by the logarithmic negativity) and possibly with finite energy. A similar problem, but using a different entanglement measure and without focusing on the energy requirement, has been considered in \cite{Pirandola2015,Wilde2017}, with fundamental implications for quantum communication.
If we apply the channel $(\tau,y)$ to one mode of a TMSS, then, in the limit of infinite squeezing, we get the Choi state \cite{Giedke2002,Serafini2017} associated to the channel. Its entanglement can be computed \cite{Niset2009} giving:
%\begin{align} \label{Echoi}
$E_\mathcal N^{({\rm Choi})}=\max\{0, -\log [y/(1+\tau)]\}$.
%\end{align}
Consider a resource state identified by $V_{AB}$ having finite entanglement $E_\mathcal N=2 r$, with $r \ge 0$. Since the total entanglement shared between Alice and Bob cannot increase under any teleportation process, one must necessarily have  $2r \ge E_\mathcal N^{({\rm Choi})}$ \cite{Mari2008,Niset2009,Pirandola2015}.  The latter bound
%, combined with Eq.~\eqref{Echoi},
defines the accessible region (complementary to the  gray area in Fig.~\ref{fig}) in the  space of phase-insensitive channels:
\begin{equation}\label{boundgray}
y \ge e^{-2r}  (1+\tau).
\end{equation}
This, intersected with Eq.~(\ref{tauyCPT}), identifies the region of Gaussian channels implementable with  $2r$ units of entanglement; or equivalently, the channels that, when applied locally to one mode of a two-mode system, always lead to an output with  $E_{\cal N} \leq 2r$, generalizing the entanglement-breaking condition \eqref{tauyEB} (dotted line in Fig.~\ref{fig}) which is recovered for vanishing $r$.

The bound in Eq.~(\ref{boundgray}) can be saturated by using the Choi state of the channel itself as a quantum resource \cite{Giedke2002,Pirandola2015}, with the gain set to $g=\sqrt{\tau}$. This solution however, though elegant,  is unrealistic for practical purposes because continuous variable Choi states have infinite energy. Here, instead, we find that there exist a realistic class of optimal resource states with minimum entanglement $2r=E_\mathcal N^{({\rm Choi})}$ and finite mean energy such that, with $g= \sqrt{\tau}$, they can simulate all physical channels at the boundary of the accessible region (where \eqref{boundgray} holds as an equality), with the exclusion of only two points, i.e.~the quantum limited attenuator and amplifier. These optimal resource states, in standard form \eqref{normal}, are found by fixing $c$ such that $E_\mathcal{N}=2r$, and $a$ such that \eqref{boundgray}, with $y$ given by \eqref{simul}, holds with equality:
\begin{align}
\begin{split}
&a =\frac{b+e^{-2r}(\tau-1)}{\tau}\,,
\quad c =\frac{b-e^{-2 r}}{\sqrt{\tau }}\,,    \\
&b \geq \frac{e^{2 r} \tau +e^{-2 r}-\left| \tau -1\right|}{\tau +1-e^{2 r} \left| \tau -1\right|}~, \label{coeff}
%  b\geq  \frac{\tau -e^{-2 r} \tanh r}{\tau -\tanh r}
\end{split}
\end{align}
where the condition given on $b$ is necessary to ensure that the state is physical. In this class of states we choose those with minimal mean energy $\bar{n}_{AB}=(a+b-2)/4$, given by the value of $b$ which saturates the inequality in \eqref{coeff}. These correspond to asymmetric squeezed thermal states with a unit symplectic eigenvalue and maximal $E_{\cal N}$ among all two-mode Gaussian states with the same marginals $a,b$  \cite{Adesso2005a,Adesso2004}. %Substituting Eqs.~\eqref{coeff} into \eqref{simul}, it is easy to check that a BK protocol using these resources with $g=\sqrt{\tau}$ simulates all phase-insensitive channels with noise parameter $y=e^{-2r}(1+\tau)$, for which \eqref{boundgray} holds with equality. Importantly,
Notice that the lower bound on the coefficient $b$ in \eqref{coeff} (and hence $a$) diverges only at the two extreme points: $(\tau=\tanh r,\, y=1-\tanh r)$, corresponding to the quantum limited attenuator, and $(\tau=\coth r,\, y=\coth r-1)$, corresponding to the quantum limited amplifier. For all the intermediate values of $\tau$ in the accessible range $\tanh(r) < \tau < \coth(r)$, all channels along the boundary saturating \eqref{boundgray} can be simulated using the states (\ref{coeff}) with finite mean energy $\bar{n}_{AB}$ and minimum entanglement $E_\mathcal{N}=2r$ (see Fig.~\ref{fig}).

Finally, it is natural to ask how much entanglement $r'$ is necessary instead to simulate a phase-insensitive channel $(\tau, y)$ exploiting the more familiar pure TMSS as a teleportation resource. This corresponds to fix:  $a=b=\cosh 2r'$ and $c=\sinh 2r'$ \cite{Weedbrook2012}. From Eq.~\eqref{simul} we have $g=\sqrt{\tau}$ and, for all non-entanglement-breaking values of the noise parameter $|1-\tau| \le y \le 1+\tau$, there always exist two solutions for $r'$, with
\begin{equation}\label{TMSSsimul}
r'= \frac{1}{2} \cosh^{-1}\left( \frac{y (1+\tau) -2 \sqrt{\tau [ y^2- (1-\tau)^2]}}{(1-\tau)^2}  \right),
\end{equation}
being the one corresponding to the smallest entanglement and indirectly  the smallest energy $\bar n_{AB}=\sinh^2(r' )$ required to simulate the channel via a TMSS.  Quite surprisingly, in this case the energy (and the entanglement) stays finite even for the quantum limited attenuator and amplifier, and diverges only for the identity map $(\tau=1,y=0)$. However,  simulation via TMSSs requires more entanglement compared to using the optimal states \eqref{coeff}: subsituting $y= e^{-2r}  (1+\tau)$ into \eqref{TMSSsimul}, one gets indeed $r' >r$ .

%{\color{red}QUESTO SI PUO' RIMUOVERE It's worth to notice that in order to simulate a channel such that $y=e^{-2r}(1+\tau)$ using a symmetric TMSS, the entanglement of the latter has to be
%\begin{equation}
%  r_{\rm TMSS}=\mbox{$\frac{1}{2} \log \left(\frac{e^{-2 r} \left(\sqrt{(\tau +1)^2-e^{4 r} (\tau -1)^2}+\tau +1\right)}{\left(\sqrt{\tau }-1\right)^2}\right)$}
%\end{equation}
%which is greater than $r$ for all $\tau$ s.t. $\tau\geq\tanh r$. With such a state it is also possible to implement the quantum limited attenuator, provided that its entanglement is $r_{\rm TMSS}=\frac{1}{2} \log \left(\frac{e^{-2 r} (\tanh (r)+1)}{\left(\sqrt{\tanh (r)}-1\right)^2}\right)$.}

This concludes the first part of this Letter, whose aim was to determine optimal teleportation protocols for simulating phase-insensitive Gaussian channels with finite resources. In the following we will exploit the previous results to solve an optimization problem with significant practical implications.

\emph{Optimal teleportation fidelity} ---
The success of a teleportation protocol can be quantified in terms of the fidelity between input and output states, which for a pure input $| \psi_{\rm in} \rangle$ is defined
as $\mathcal F= \langle \psi_{\rm in} | \rho_{\rm out} |\psi_{\rm in} \rangle$ \cite{Braunstein2000}. For a coherent input state $|\alpha\rangle$ the latter is proportional to the $Q$-function  \cite{Husimi1940,Gardiner1999} of the output state evaluated at the input complex amplitude $\alpha$:
$\mathcal F=  \langle \alpha|\rho_{\rm out}|\alpha\rangle= \pi Q_{\rm out}(\alpha)$.
 Here we are interested in the realistic scenario \cite{Braunstein2000,Hammerer2005} in which Alice wants to teleport an unknown input coherent state $| \alpha\rangle$, with displacement $\alpha$ sampled from a Gaussian phase space distribution
  $P_{\rm in}^\lambda(\alpha)=(\lambda/{\pi})e^{-\lambda~|\alpha|^2}$. This prior corresponds to the $P$-function \cite{Glauber1963,Gardiner1999} of a thermal ensemble with input mean energy $\bar{n}_{\rm in}$ given by the variance $\lambda^{-1}$.
The average teleportation fidelity over the input ensemble can be expressed as the integral overlap of the two functions:
\begin{equation}\label{eq:avgfid}
  \mbox{$\bar{\mathcal{F}}^\lambda=\pi \int_{\mathbb{C}}d^2\alpha~P_{\rm in}^\lambda(\alpha) Q_{\rm out}(\alpha)$}\,.
\end{equation}

 Now let us consider that Alice and Bob share a two-mode resource state with fixed entanglement $E_\mathcal N=2r$. We want to optimize the average fidelity over all possible Gaussian phase-insensitive teleportation schemes. This task appears {\it prima facie} quite complex;  however, thanks to the previous analysis, we can limit the optimization over the two-parameter space $(\tau,y)$ of accessible phase-insensitive Gaussian channels with the entanglement constraint \eqref{boundgray}, without delving into the specifics of the teleportation protocol. %We have shown in fact that for each accessible channel there exists at least one protocol that can implement it with the supplied entanglement.
The action of any such channel
%a phase-insensitive Gaussian channel parametrized by $(\tau,y)$
on a coherent input state produces, according to Eqs.~\eqref{GChannel} and \eqref{pic}, a thermal output state with displacement $d_{\rm out}=(\!\sqrt{2 \tau}\,\mbox{Re}\,\alpha,\sqrt{2 \tau}\,\mbox{Im}\,\alpha)^\top$ and covariance matrix $V_{\rm out}= (y+\tau) \mathbb{1}$. The corresponding $Q$-function, evaluated at the input phase space point  $\alpha$, is  $Q_{\rm out}^\lambda(\alpha)=[2 e^{-2 (1-\sqrt{\tau})^2 |\alpha |^2/(y+ \tau +1)}]/[\pi (y+ \tau +1) ]$.
Substituting this into Eq.~\eqref{eq:avgfid}, %and performing a simple Gaussian integral,
we get the average teleportation fidelity
\begin{equation}\label{eq:fidelity}
  \bar{\mathcal{F}}^\lambda(\tau,y)={2\lambda}/\left[{2(1-\sqrt{\tau})^2+\lambda(1+y+\tau)}\right]~.
\end{equation}
The previous expression depends nontrivially on the transmissivity parameter $\tau$ while, as expected, it is monotonically decreasing with  the noise parameter $y$ (see Fig.~\ref{fig} for a contour plot).
For a fixed entanglement $E_\mathcal N=2r$, $y$ is lower bounded by \eqref{boundgray} and so the maximum $\bar{\mathcal{F}}^\lambda$ must be on the line $y=e^{-2r}(1+\tau)$, delimiting the set of implementable channels from the unaccessible region (gray area in Fig.~\ref{fig}).  Inserting this into \eqref{eq:fidelity}, and optimizing with respect to $\tau$ within the completely positive region \eqref{tauyCPT}, we get:
$\tau_{\rm opt}=\max \left\{  {\tanh r}\,, \frac{e^{2 r}}{(e^r + \lambda \cosh r )^2}\right\}$. The
corresponding optimal average fidelity is finally:
 \begin{equation} \label{Fopt}
  \bar{\mathcal{F}}^\lambda_{{\rm opt}}(r) = \left\{
                                               \begin{array}{ll}
                                                 \displaystyle\frac{\lambda }{\lambda +\big(1 - \sqrt{\tanh r}\big)^2},\  & \tanh r \geq \frac{e^{2 r}}{(e^r + \lambda \cosh r )^2};\\[-.2cm] & \\
                                                 \displaystyle\frac{e^r (1+\lambda +\tanh r )}{2 e^r+\lambda \ \cosh r},\  & \hbox{otherwise}.
                                               \end{array}
                                             \right.
\end{equation}
The associated optimal teleportation channel, denoted by a triangle in Fig.~\ref{fig}, can be simulated via a non-unit gain BK protocol based on the class of resource states of  Eq.~\eqref{coeff}.  However, when the first case of Eq.~(\ref{Fopt}) holds, the optimal channel is a quantum limited attenuator for which the needed energy diverges as previously discussed. When instead the second case holds, which happens for sufficiently large input variance $\lambda^{-1}$,
 %or for sufficiently small resource entanglement $r$,
 the optimal channel can be implemented  with finite energy.

\emph{Discussion} --- We now discuss some implications of our general formula for the optimal average fidelity given in Eq.~\eqref{Fopt}. In particular we are going to recover, as special cases,  a number of results obtained in previous literature.

The first special case that we consider is $r=0$ (see diamond in Fig.~\ref{fig}), that is, no shared entanglement, for which we get:
\begin{equation}\label{Fclass}
\tau_{\rm opt}=(1+\lambda)^{-1}, \quad y_{\rm opt}=1+\tau_{\rm opt}, \quad  \mathcal{F}^\lambda_{{\rm opt}}(0)=\frac{1+\lambda}{2+\lambda},
\end{equation}
yielding the maximum fidelity achievable with any measure-and-prepare strategy, i.e., the classical benchmark derived in \cite{Hammerer2005}. Moreover, for $r=0$ the two-mode state \eqref{coeff} reduces to the vacuum and
the BK protocol is equivalent to a heterodyne detection at Alice's site, followed by a repreparation of a coherent state at Bob's site with  $g=(1+\lambda)^{-1}$. This is exactly the optimal ``cheating'' strategy originally proposed in \cite{Braunstein2000}.

The second special case is when $\lambda \rightarrow 0$, corresponding to the teleportation of uniformly distributed coherent states (i.e., with unbounded input mean energy). In this limit we get:
\begin{equation} \label{FBK}
\tau_{\rm opt}=1, \quad y_{\rm opt}=2 e^{-2r}, \quad  \bar{\mathcal{F}}^0_{{\rm opt}}(r)=\big(1+e^{-2r}\big)^{-1},
\end{equation}
consistently with the results obtained in Refs.~\cite{Adesso2005,Mari2008}. Moreover for $\lambda \rightarrow 0$  the optimal resource \eqref{coeff} reduces to a pure TMSS, recovering the standard BK scheme with unit gain \cite{Braunstein1998}.

In general, for arbitrary values of $\lambda$ and $r$, one may compare the optimal teleportation strategy derived in this Letter [Eq.~\eqref{Fopt}] with a conventional BK scheme based on a pure TMSS and optimized gain $g$ \cite{Furusawa1998,Polkinghorne1999,Ide2002,Hofmann2000}. Using Eqs.~\eqref{simul} and optimizing Eq.~\eqref{eq:fidelity} with respect to $\tau=g^2$, we get: $g_{\rm opt}=(2 + \lambda \sinh 2 r)/(2 + \lambda + \lambda \cosh 2 r )$, and
\begin{align}\label{FTMSS}
\bar{\mathcal{F}}^\lambda_{\mbox{\tiny \rm TMSS}}(r)=(\text{sech}^2r+\lambda)/(2+\lambda -2 \tanh r),
\end{align}
corresponding to the circle in Fig.~\ref{fig}. One sees that the teleportation based on a TMSS never achieves the optimal Gaussian strategy (triangle) for any $\lambda>0$, despite approximating it well and beating the classical benchmark. Precisely, one has:
%\begin{align}
 $ \bar{\mathcal{F}}^\lambda_{{\rm opt}} (r) \ge \bar{\mathcal{F}}^\lambda_{\mbox{\tiny \rm TMSS}}(r)\ge \bar{\mathcal{F}}^\lambda_{{\rm opt}} (0)$,
%\end{align}
where the first inequality is saturated for $\lambda \rightarrow 0$ or $r=0$, while the second one only for $r=0$.
Therefore, if one takes into account only the shared entanglement, pure TMSSs are suboptimal for teleporting coherent states with non-uniform distribution. On the other hand, teleportation with a TMSS may still represent an experimentally practical solution, e.g.~when  $\tanh r\geq e^{2r}/(e^r+\lambda\cosh r)^2$  [see Eq.~\eqref{Fopt}], in which case the energy of the optimal state \eqref{coeff} diverges while it is finite for a TMSS.

%{\color{red}We also want to emphasize that even though $\tanh r\leq e^{2r}/(e^r+\lambda\cosh r)^2$, i.e. when the optimal state \eqref{coeff} does not exist with finite mean energy, it is possible also in this case to construct an asymmetric squeezed thermal state with a unit symplectic eigenvalue whose performance will always be better than the TMSS and can be arbitrarily close to the optimal one, when energetic enough:
%\begin{align}
% a &=b \coth r-\cosh (2 r) (\coth r-1)\,\nonumber \\
% c&=e^{-2 r} \sqrt{\left(a e^{2 r}-1\right) \left(b e^{2 r}-1\right)}\,,  \nonumber \\
%  b&\geq\cosh(2r)~, \label{coeff2}
%\end{align}
%and appropriate choice of the gain (we don't report here its cunbersone expression). When the inequality in \eqref{coeff2} is saturated we recover the TMSS state.
%}

\emph{Conclusions} ---
In this Letter we determined a class of realistic continuous variable teleportation protocols that are optimal for simulating phase-insensitive Gaussian channels employing a minimum amount of shared entanglement. Excluding the pathological cases of the quantum limited attenuator and amplifier, our teleportation schemes rely on feasible resource states with a finite mean energy and are thus quite appealing for practical applications in quantum communication \cite{Braunstein2005,Pirandola2015,Wilde2017,Wildenew,Pirlanew}.

By exploiting such effective equivalence between Gaussian channels and teleportation schemes, we considered the relevant task of teleporting an alphabet of single-mode coherent states sampled from a Gaussian phase space distribution with finite variance \cite{Braunstein2000}, via a two-mode Gaussian resource with finite entanglement \cite{Adesso2004}. We determined the optimal average fidelity analytically, solving a longstanding open problem in continuous variable quantum teleportation \cite{Pirandola2006,Weedbrook2012}, and recovering many previous results as special cases \cite{Braunstein2000,Hammerer2005,Adesso2005,Mari2008}.

Future generalizations of our analysis may include the identification of finite energy resource states and effective teleportation protocols for simulating phase-sensitive and possibly multimode Gaussian channels.
%This could allow us to optimize the teleportation of coherent or squeezed input states sampled with asymmetric prior distributions.
Another possible direction concerns the optimal use of steering \cite{Wiseman2007}, rather than entanglement, as a resource for secure teleportation \cite{He2015,Kogias2014,Pietro2017}.
%}, and the determination of the optimal average fidelity given a fixed degree of Gaussian steering  in the shared resource state.

\begin{acknowledgments}
\emph{Acknowledgments} ---
We thank Marco Cianciaruso for fruitful discussions in the early stages of this project, and Stefano Pirandola, Cosmo Lupo, Riccardo Laurenza, Mark Wilde, and Eneet Kaur for further discussions after our paper first appeared. G.A.~thanks Pamela Ott for inspiring exchanges about entanglement and its artistic depiction. We acknowledge financial support from the European Research Council (ERC) under the Starting Grant GQCOP (Grant No.~637352) and the Foundational Questions Institute under the Physics of the Observer Programme (Grant No.~FQXi-RFP-1601).
\end{acknowledgments}

%\bibliographystyle{apsrevfixedwithtitles}
%\bibliography{biblioteleport}

%merlin.mbs apsrev4-1.bst 2010-07-25 4.21a (PWD, AO, DPC) hacked
%Control: key (0)
%Control: author (72) initials jnrlst
%Control: editor formatted (1) identically to author
%Control: production of article title (1) required
%Control: page (0) single
%Control: year (1) truncated
%Control: production of eprint (0) enabled
%

\end{document}